Nondiffusive thermal transport from micro/nanoscale sources producing nonthermal phonon populations exceeds Fourier heat conduction

Vazrik Chiloyan[a], Samuel Huberman[a], Alexei A. Maznev[b], Keith A. Nelson[b], Gang Chen[a]*

[a]Department of Mechanical Engineering, Massachusetts Institute of Technology, Cambridge, Massachusetts 02139, USA
[b]Department of Chemistry, Massachusetts Institute of Technology, Cambridge, Massachusetts 02139, USA

We study nondiffusive thermal transport by phonons at small distances within the framework of the Boltzmann transport equation (BTE) and demonstrate that the transport is significantly affected by the distribution of phonons emitted by the source. We discuss analytical solutions of the steady-state BTE for a source with a sinusoidal spatial profile, as well as for a three-dimensional Gaussian "hot spot," and provide numerical results for single crystal silicon at room temperature. If a micro/nanoscale heat source produces a thermal phonon distribution, it gets hotter than predicted by the heat diffusion equation; however, if the source predominantly produces low-frequency acoustic phonons with long mean free paths, it may get significantly cooler than predicted by the heat equation, yielding an enhanced heat transport.

*Corresponding author: gchen2@mit.edu

Thermal transport in semiconductors is dominated by phonons. At room temperature (RT) and on the macroscale, phonon-mediated heat transport conforms to Fourier's law and the heat conduction equation. However, when the characteristic distance of the heat transfer comparable to the mean free path (MFP) of heat-carrying phonons, thermal transport no longer follows the predictions of the heat diffusion equation [1–7]. At room temperature (RT) the phonon MFP is short, but recent theoretical and experimental studies indicated that low-frequency acoustic phonons with MFPs in the micron range still contribute significantly to thermal conductivity of single crystal materials such as silicon [6,8–10]. Consequently, non-diffusive heat transport at RT is observed on micrometer and sub-micrometer scales [4,6,7,10,11]. It has been observed that the heat transport from a small heat source is reduced compared to the prediction of the heat equation because the contribution of long-MFP phonons is smaller than the diffusion model predicts [3–7]. Consequently, a micro/nanoscale "hot spot" will be hotter than expected based on the heat equation, which obviously has significant implications for thermal management of microelectronic devices [12–14].

Thus far in the analysis of nondiffusive phonon transport with the Boltzmann transport equation (BTE), it was assumed that heat sources produce a phonon distribution corresponding to the thermal equilibrium at a given temperature [15–17]. However, the spectrum of emitted phonons is expected to depend on the physical nature of the heat source.
For example, many experiments aimed at understanding nondiffusive thermal transport use optical excitation [6,10,18,19], in which case the phonon distribution generated depends on how photoexcited carriers transfer energy to the lattice and will in general not be a thermal distribution. The phonon distribution produced by ohmic heating in a semiconductor device will also depend on the details of electron-phonon scattering In this report, we go beyond the "thermal" source model and explore nondiffusive heat transport from nonthermal sources.

We will show that the thermal transport by phonons in the nondiffusive regime is strongly dependent on the distribution of phonons generated by the heating source. While for a thermal distribution, the decrease of the size of the heat source yields reduced transport and higher temperatures than predicted by Fourier's law, this is not necessarily the case for nonthermal distributions. We will demonstrate that for a source which predominantly generates long MFP phonons, the thermal transport predicted by the BTE is enhanced and may significantly exceed the prediction by the heat equation, yielding an effective thermal conductivity larger than the regular macroscale conductivity.

We start with a study of a model problem with a steady state spatially sinusoidal heat source generating a steady-state "thermal grating" (TG). We will analyze the analytical solution of the BTE for this simple case and present numerical results for different phonon source distributions for silicon at room temperature. An arbitrary steady-state heat source can be represented by a superposition of such thermal gratings. As an example, we will study a three-dimensional Gaussian source mimicking a "hot spot" that could be produced by ohmic heating inside a microelectronic device.

We consider thermal transport in silicon at room temperature and start with the isotropic BTE under the relaxation time approximation (RTA) [20]:

$$\frac{\partial g_\omega}{\partial t} + \vec{v}_\omega \cdot \vec{\nabla} g_\omega = \frac{g_0 - g_\omega}{\tau_\omega} + \frac{Q_\omega}{4\pi} \tag{1}$$

The RTA is valid for silicon around room temperature, and has been used to accurately predict its thermal conductivity [8,9,21]. In Eq. (1), $g_\omega$ is the phonon energy density per unit frequency interval per unit solid angle above the equilibrium spectral energy density at background temperature $T_0$, related to the phonon distribution function $f_\omega$ and density of states $D(\omega)$ as $g_\omega = \frac{\hbar\omega D(\omega)}{4\pi}(f_\omega - f_0(T_0))$. $\vec{v}_\omega$ is the group velocity, $\tau_\omega$ is the relaxation time, and $g_0$ is the equilibrium energy density, given by $g_0 \approx \frac{1}{4\pi} C_\omega \Delta T$ in the linear response regime where we assume the temperature rise due to the heating is small compared to the background room temperature. The spectral heat capacity is given by $C_\omega = \hbar\omega D(\omega) \frac{df_0}{dT_0}$, and the equilibrium distribution is given by Bose-Einstein statistics as $f_0(T_0) = \frac{1}{\exp\left(\frac{\hbar\omega}{k_B T_0}\right) - 1}$. We utilize the subscript $\omega$ as a short hand notation to denote a phonon of a given branch and frequency. The energy density in the BTE is obtained by integrating the nonequilibrium energy density of phonons over the solid angle and frequency modes and branches, i.e. $U = \int d\omega \int d\Omega\, g_\omega$.

The spectral volumetric heat generation term can be written as $Q_\omega = p_\omega Q$, where $Q$ is the macroscopic volumetric heat generation rate, and $p_\omega$ represents the degree to which a phonon mode is excited by the heating. These values are normalized such that the sum over the branches and phonon frequencies add up to unity, i.e. $\int d\omega p_\omega = 1$. The spatial distribution of $Q_\omega$ is given by the macroscopic heat generation rate $Q$, while $p_\omega$ only depends on the physics of the heating process.

In the steady state TG geometry, the volumetric heat generation rate is given by the functional form $Q = \tilde{Q}e^{i\vec{q}\cdot\vec{r}}$, where $q$ is the grating wavevector, whose magnitude $q$ is given in terms of the grating period as $q = 2\pi/\lambda$, and $\tilde{Q}$ is the amplitude of the volumetric heat generation rate. The BTE solution will also have a spatial sinusoidal profile, yielding:

$$g_\omega = e^{i\vec{q}\cdot\vec{r}} \frac{1}{4\pi} \frac{1}{1+i\vec{q}\cdot\vec{v}_\omega \tau_\omega} \left[ C_\omega \Delta\tilde{T} + p_\omega \tau_\omega \tilde{Q} \right] \qquad (2)$$

Summing over the solid angle and phonon spectrum yields the energy density as a functional of the temperature and heat generation rate:

$$U = \Delta\tilde{T} e^{i\vec{q}\cdot\vec{r}} \int d\omega C_\omega \frac{\arctan(q\Lambda_\omega)}{q\Lambda_\omega} + \tilde{Q}e^{i\vec{q}\cdot\vec{r}} \int d\omega p_\omega \tau_\omega \frac{\arctan(q\Lambda_\omega)}{q\Lambda_\omega} \qquad (3)$$

We close the problem by utilizing the energy conservation equation, defining the pseudo temperature $\Delta T$:

$$\int d\omega \int d\Omega \frac{1}{\tau_\omega}\left[ \frac{C_\omega}{4\pi}\Delta T - g_\omega \right] = 0 \qquad (4)$$

It should be noted that for a nonequilibrium distribution, the temperature is not strictly defined. While the temperature of Eq. (4) is used within the framework of the RTA [16,22,23], it is also common to define the temperature as the ratio of the energy density from the BTE to the heat capacity, $U/C$ [1,24,25]. To avoid any ambiguity in defining the temperature, we will present our final results in terms of the energy density rather than the temperature.

Inputting the solution to the BTE into the energy conservation equation yields the pseudo temperature:

$$\Delta T = \tilde{Q} e^{i\vec{q}\cdot\vec{r}} \frac{\int d\omega p_\omega \frac{\arctan(q\Lambda_\omega)}{q\Lambda_\omega}}{\int d\omega \frac{C_\omega}{\tau_\omega}\left[1 - \frac{\arctan(q\Lambda_\omega)}{q\Lambda_\omega}\right]} \qquad (5)$$

The arctan functions appear due to the isotropic approximation and arise from the solid angle integrals performed in the energy conservation equation. Another consequence of the isotropic assumption is that the equation only depends on the thermal grating wavevector through its magnitude $q$. Equation (5) is the steady state version of the results obtained previously [16]. Substituting the pseudo temperature of Eq. (5) into Eq. (3) yields the energy density:

$$U_{BTE} = \tilde{Q} e^{i\vec{q}\cdot\vec{r}} \left\{ \int d\omega p_\omega \tau_\omega \frac{\arctan(q\Lambda_\omega)}{q\Lambda_\omega} + \frac{\left[\int d\omega p_\omega \frac{\arctan(q\Lambda_\omega)}{q\Lambda_\omega}\right]\left[\int d\omega C_\omega \frac{\arctan(q\Lambda_\omega)}{q\Lambda_\omega}\right]}{\int d\omega \frac{C_\omega}{\tau_\omega}\left[1 - \frac{\arctan(q\Lambda_\omega)}{q\Lambda_\omega}\right]} \right\} \qquad (6)$$

In the diffusive limit, where $q\Lambda_\omega \to 0$, the second term dominates, and a Taylor expansion of the arctan functions in Eq. (6) yields a result coinciding with the solution of the Fourier heat conduction equation:

$$U_{Fourier} = \frac{\tilde{Q}}{\alpha q^2} e^{i\vec{q}\cdot\vec{r}} \qquad (7)$$

with a thermal diffusivity given by $\alpha = \frac{1}{3C}\int d\omega C_\omega v_\omega \Lambda_\omega$.

In the ballistic limit, where $q\Lambda_\omega \to \infty$, the first term dominates in the expression for the energy density of Eq. (6), and the energy density simplifies to:

$$\lim_{q\Lambda_\omega \to \infty} U_{\text{BTE}} = \tilde{Q} e^{i\vec{q}\cdot\vec{r}} \frac{\pi}{2} \frac{1}{q} \int d\omega \frac{p_\omega}{v_\omega} \tag{8}$$

The trend in the ballistic limit is that the energy density is proportional to the inverse grating wavevector, i.e. proportional to the grating period. By contract, the Fourier energy density of Eq. (7) is proportional to the square of the grating period.

We shall now consider three kinds of the source phonon distribution: (i) a "thermal" source producing a thermal phonon distribution; (ii) a source only generating optical phonons (for example, the recombination of photoexcited carriers will predominantly generate optical phonons); (iii) a high MFP filter source only generating phonons with MFPs exceeding a certain threshold $\Lambda$. The corresponding distribution functions (all properly normalized) are given by:

$$p_\omega^{\text{thermal}} = \frac{C_\omega}{C}$$

$$p_\omega^{\text{optical}} = \begin{cases} 0 & \omega \in \text{acoustic branch} \\ \dfrac{C_\omega}{C_{\text{optical}}} & \omega \in \text{optical branch} \end{cases} \tag{9}$$

$$p_\omega^{\text{filter}} = \frac{C_\omega \Theta(\Lambda_\omega - \Lambda)}{\int d\omega' C_{\omega'} \Theta(\Lambda_{\omega'} - \Lambda)}$$

The study of the high MFP filter source distribution is motivated by the aim to minimize the energy density, and thus maximize the thermal transport. Minimizing the energy density requires minimizing the entire expression of Eq. (6). As the first term dominates in the ballistic limit, and the second term dominates in the diffusive limit, the optimal distribution that minimizes the energy density will be different for different grating periods. One can see that the use of a high MFP filter reduces the second term in Eq. (6). Our expectation is that it will also reduce the total energy density in the weakly non-diffusive regime when the second term in Eq. (6) still dominates; as we will see shortly, this is indeed confirmed by numerical calculations.

Figure 1a shows the ratio of the energy density predicted by the BTE from Eq. (6) to the Fourier heat conduction energy density of Eq. (7) as a function of the grating period $\lambda$ for silicon, with material properties from DFT utilized previously [22]. As expected, at very large grating periods, the ratio approaches unity, since the prediction from the BTE will be the same as the Fourier heat conduction equation in the diffusive limit. The difference between thermal and optical distributions is small for grating periods on the order of 1 micron and higher. The difference grows at grating periods smaller than 1 micron, where the optical distribution would yield a 50% larger energy density than the thermal distribution at a grating of 100 nm. Thus, the importance of the source distribution is greater as the length scales of the experiment get smaller.

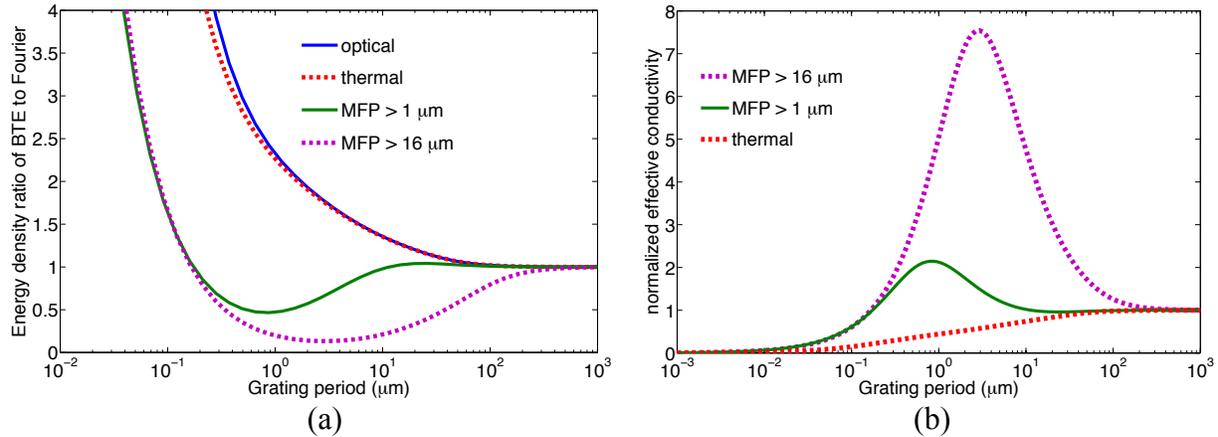

FIG. 1. Steady state thermal grating geometry comparison between the BTE and Fourier heat conduction equation. (a) The energy density ratio of BTE to Fourier's law and (b) its inverse which yields the effective conductivity show the comparison among the different source phonon distributions.

The high MFP filter achieves thermal transport that yields smaller energy densities than the thermal distribution, and even smaller than those predicted by the Fourier heat conduction equation. According to Eq. (7), the thermal conductivity in the diffusive transport regime is inversely proportional to the energy density, $k_{Fourier} = \frac{\tilde{Q}C}{q^2 U_{Fourier}} e^{i\vec{q}\cdot\vec{r}}$. Thus by matching the results of the BTE and the Fourier heat conduction equation, we can introduce an effective thermal conductivity for nondiffusive heat transport in a steady state TG, $k_{eff} = \frac{\tilde{Q}C}{q^2 U_{BTE}} e^{i\vec{q}\cdot\vec{r}}$. The ratio of the effective conductivity to the regular Fourier conductivity, given by $k_{eff}/k_{Fourier} = U_{Fourier}/U_{BTE}$, is shown in Fig. 1b. Note that we use the term effective thermal conductivity despite the fact that Fourier's law is invalid in the nondiffusive transport regime. Our purpose here is to quantify the heat dissipation efficiency compared to the Fourier heat conduction equation: if $k_{eff}/k_{Fourier} = 2$, for example, then it means that the energy density is one half of what the Fourier heat conduction predicts. As can be seen from Fig. 1b, a source generating phonons with the highest MFPs of silicon, with approximately 16 $\mu$m MFP, results in an effective thermal conductivity that is over 7 times larger than the regular Fourier thermal conductivity at a TG period of 3 $\mu$m. The material properties of silicon were obtained from DFT, similar to the work from Ref. [22]. Compared to the thermal or optical distributions, the high MFP filter source results in an energy density that is more than an order of magnitude smaller. This extraordinary enhancement of thermal transport, beyond even that of Fourier heat conduction, indicates the importance of the source phonon distribution and shows that size effects from a heating length scale should not be thought of as independent of the source distribution. Depending on the distribution, the thermal transport can be reduced or enhanced compared to the predictions of Fourier's law.

We noted that in the ballistic limit, given the linear scaling with respect to the grating period compared to the quadratic scaling given by Fourier's law, for small enough grating periods, the energy density of the BTE will be higher than the result from the Fourier heat conduction equation. This is exhibited in Fig. 1a for grating periods shorter than 200nm. Therefore, the reduction of the energy density (i.e., the enhancement of the thermal transport) relative to the Fourier heat conduction equation is only possible within an intermediate range of TG periods.

The thermal grating provides a simple one-dimensional geometry with which we could quantify the effect of the various phonon source distributions. Furthermore, an arbitrary steady state source can be represented as a superposition of gratings via the Fourier transform. As an example, we now consider a three-dimensional steady state hot spot, which can be produced, for example, by ohmic heating inside a microelectronic device. We consider a heat source described by a radial Gaussian, given by $Q = \bar{Q} \frac{1}{R^3 (2\pi)^{\frac{3}{2}}} \exp\left(-\frac{r^2}{2R^2}\right)$ in 3D space. $R$ defines the size of the hot spot, and $\bar{Q}$ represents the power being deposited into the system. We will use the energy density at the center of the hot spot, i.e. $r = 0$ to compare the predictions of the BTE and Fourier heat conduction equations. Utilizing the spatial Fourier transform of the radial Gaussian, the heat generation rate can be written as a superposition of 1D thermal gratings as $Q = \frac{1}{(2\pi)^3} \int e^{i\vec{q}\cdot\vec{r}} \bar{Q} \exp\left(-\frac{q^2 R^2}{2}\right) d^3q$. Taking a superposition of the one-dimensional grating solutions of Eq. (6) and Eq. (7), weighted by the coefficients of the Fourier transform for the radial Gaussian, we obtain the energy density at the center of the Gaussian hot spot:

$$U_{\text{BTE}}(r=0) = \bar{Q}\frac{1}{2\pi^2}\frac{1}{R^3}\int_0^\infty dt \exp\left(-\frac{t^2}{2}\right)t^2 \left\{ \frac{\int d\omega p_\omega \tau_\omega \frac{\arctan(t\eta_\omega)}{t\eta_\omega} + \left[\int d\omega p_\omega \frac{\arctan(t\eta_\omega)}{t\eta_\omega}\right]\left[\int d\omega C_\omega \frac{\arctan(t\eta_\omega)}{t\eta_\omega}\right]}{\int d\omega \frac{C_\omega}{\tau_\omega}\left[1 - \frac{\arctan(t\eta_\omega)}{t\eta_\omega}\right]} \right\} \quad (10)$$

$$U_{\text{Fourier}}(r=0) = \bar{Q}\frac{1}{\alpha R}\frac{1}{(2\pi)^{\frac{3}{2}}}$$

The isotropy of the system allowed us to analytically compute the integral over the solid angle, reducing the 3D inverse Fourier transform integral into an integral over the radial variable. We defined the nondimensional variables $t = qR$ and $\eta_\omega = \Lambda_\omega / R$ to simplify the expression.

Figure 2a shows the energy density from the Fourier heat conduction equation as well as from the BTE for several source distributions. The energy density is taken at the center of the hot spot as a function of the size of the hot spot $R$, with a given input power of $\bar{Q} = 1\,\mu\text{W}$. While the energy density at the center increases with decreasing spot size as the energy is more concentrated at the center, for spot sizes near 300 nanometers, the high MFP filter distribution with the 16 micron threshold yields an energy density an order of magnitude lower than that of the thermal distribution, and 5 times lower than predicted by the heat equation.

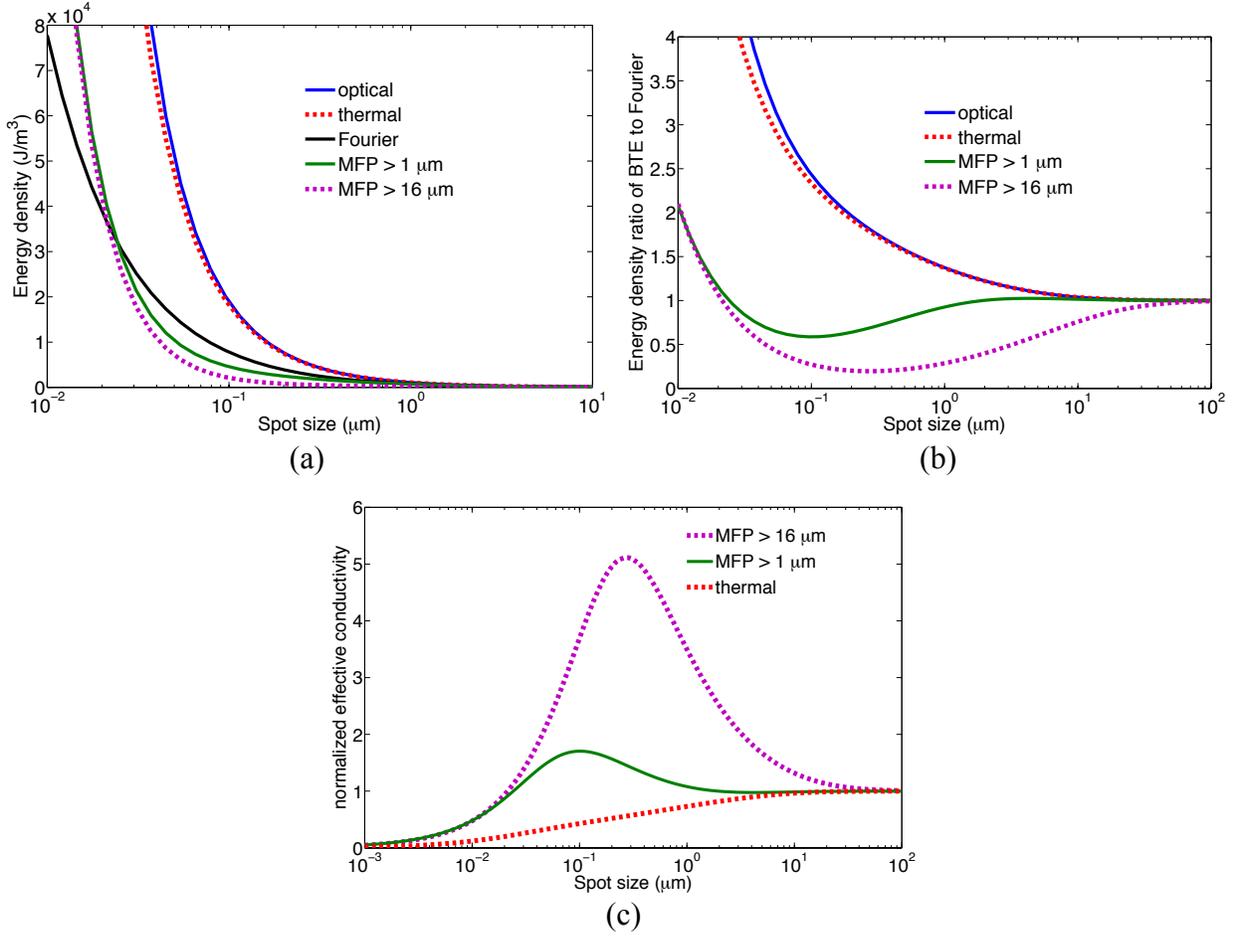

Fig. 2. Heat transport from a 3D Gaussian hot spot for different phonon source distributions vs. the size of the hot spot. (a) The energy density for a given input power, (b) the energy density relative to that given by Fourier's law, and (c) the effective conductivity relative to that given by Fourier's law.

By matching the energy densities at the center of the hot spot $r = 0$, we define a normalized effective thermal conductivity for the 3D Gaussian hot spot, obtained by taking the ratio of the Fourier energy density to the BTE energy density from Eq. (10), i.e. $\kappa_{\text{hot spot}} \equiv U_{\text{Fourier}}(r=0)/U_{\text{BTE}}(r=0)$. We see a factor of 5 times enhancement in the effective thermal conductivity at 300nm in Fig. 2c. The high MFP distribution predicts an energy density that is lower than the prediction from the Fourier heat conduction equation for spot sizes of approximately 20nm and larger.

In conclusion, we have investigated the effect of the phonon distribution produced by a small heat source on nondiffusive thermal transport. By comparing the energy density in response to volumetric heating from the BTE and from the Fourier heat conduction equation for a 1D thermal grating as well as for a 3D Gaussian hot spot, we have dispelled the notion that the size effect always leads to a reduction in thermal transport compared to the prediction of Fourier's law. Depending on the source phonon distribution, the size effect can reduce or enhance thermal transport; in the latter case, the effective micro/nanoscale thermal conductivity can become larger than the regular macroscale conductivity. We believe that a source predominantly generating long-MFP phonons and thus yielding enhanced thermal transport could be realized in practice. For example, in ohmic heating of a semiconductor device, the distribution of phonons excited is determined by the electron-phonon interaction. Intraband electron-phonon scattering, which typically plays a dominant role in the resistivity of semiconductors at room temperature, involves small-wavevector phonons at the center of the Brillouin zone, which will include long-MFP acoustic phonons [26]. If the predominant emission of such long-MFP phonons could enhance thermal transport from nanoscale hot spots, this would have significant implications for the thermal management of microelectronic devices.

This work was supported by S³TEC, an Energy Frontier Research Center funded by the U.S. Department of Energy, Office of Basic Energy Sciences, under Award No. DE- SC0001299/DE-FG02-09ER46577.

*Classic Texts in the Physical Sciences)* (Oxford University Press, London, 1960).
[21] D. A. Broido, M. Malorny, G. Birner, N. Mingo, and D. A. Stewart, Appl. Phys. Lett. **91**, 231922 (2007).
[22] K. C. Collins, A. A. Maznev, Z. Tian, K. Esfarjani, K. A. Nelson, and G. Chen, J. Appl. Phys. **114**, 104302 (2013).
[23] V. Chiloyan, L. Zeng, S. Huberman, A. A. A. Maznev, K. A. K. A. Nelson, and G. Chen, J. Appl. Phys. **120**, 25103 (2016).
[24] Q. Hao, G. Chen, and M.-S. S. Jeng, J. Appl. Phys. **106**, 114321 (2009).
[25] J.-P. M. Péraud and N. G. Hadjiconstantinou, Phys. Rev. B **84**, 205331 (2011).
[26] B. Liao, B. Qiu, J. Zhou, S. Huberman, K. Esfarjani, and G. Chen, Phys. Rev. Lett. **114**, 115901 (2015).
15